\documentclass[journal]{IEEEtran}

\usepackage{graphicx}
\usepackage{url}
\usepackage[cmex10]{amsmath}
\usepackage{microtype}

\begin{document}


\title{\LARGE{A baseline-free method to identify responsive customers\\ on dynamic time-of-use tariffs}}
\author{James R. Schofield,~\IEEEmembership{Member,~IEEE,}
        Simon H. Tindemans,~\IEEEmembership{Member,~IEEE,}
        and~Goran~Strbac,~\IEEEmembership{Member,~IEEE}
\thanks{The authors are with the Department of Electrical and Electronic Engineering, Imperial College London, London SW7 2AZ, UK. \emph{(corresponding e-mail: s.tindemans@imperial.ac.uk}) This research was supported by the \emph{Low Carbon London} project, which was funded through the \emph{Low Carbon Networks Fund} programme, administered by the UK Regulator, \emph{Ofgem}.}%
}

\maketitle

\begin{abstract}
Dynamic time-of-use tariffs incentivise changes in electricity consumption. This paper presents a non-parametric method to retrospectively analyse consumption data and quantify the significance of a customer's observed response to a dynamic price signal without constructing a baseline demand model. If data from a control group is available, this can be used to infer customer responsiveness---individually and collectively---on an absolute scale. The results are illustrated using data from the Low Carbon London project, which included the UK's first dynamic time-of-use pricing trial. 
\end{abstract}

\begin{IEEEkeywords}
demand response, dynamic pricing, nonparametric statistics, power demand, smart meters.
\end{IEEEkeywords}


\section{Introduction}
\IEEEPARstart{D}{ynamic} time-of-use (dToU) tariffs are characterised by time-dependent electricity prices without a predetermined pattern. They are an important tool to elicit demand response from residential and small business loads. 
Methods to quantify the impact of dToU price signals have to date focused on the demand baseline~\cite{Mathieu2011,Wijaya2014}; the predicted demand in the hypothetical absence of a price signal. 
This paper proposes an elegant method for retrospective analysis of customer demand data that establishes a signal-to-noise metric for responsiveness to dToU tariffs. Notably, the method is nonparametric and does not depend on demand baselines or behavioural models. The method is illustrated using one year~(2013) of half-hourly metered consumption data from the Low Carbon London (LCL) dToU trial~\cite{Schofield2015}. After data cleansing a total of 4,756 smart metered households were available, of which 988 opted to receive the experimental dToU tariff. 

\section{Description of the method}
Consider customers $i=1,\ldots,N$, whose electricity consumption has been recorded in the form of time series $(c^i_1, \ldots, c^i_T)$ (e.g. half-hourly consumption data). The customers have received a dynamic time-of-use price signal $(p_1, \ldots, p_T)$ corresponding to the elements of the consumption time series. The electricity bill for customer $i$ is therefore
\begin{equation}
b^i = {\sum}_{t=1}^T p_t c_t^i.
\end{equation}

We aim to quantify our confidence that the recorded consumption data $c^i_t$ has been influenced by the price signal $p_t$. Let $\pi:\{1,\ldots,T\} \rightarrow \{1,\ldots,T\}$ be a random permutation that shuffles days (so as to maintain the structure of price events) in the price signal and define a randomised electricity bill (a random variable) as
\begin{equation}
B^i = {\sum}_{t=1}^T p_{\pi(t)} c^i_t. \label{eq:randombilldef}
\end{equation}

We postulate that a customer $i$ who consumes electricity \emph{without regard} for the price signal receives an actual bill $b^i$ that is in line with the distribution of randomised bills $B^i$. If, on the other hand, a customer deliberately responds to tariff signals, the realised bill $b^i$ should be lower than most randomised bills. A responsiveness metric can thus be defined as the fraction
\begin{equation} \label{eq:qDefinition}
\phi^i = \mathrm{Pr}(B^i > b^i), \qquad \phi^i \in [0,1].
\end{equation}
It is a measure of conformity between a household's energy consumption and the tariff signal, with high values ($\approx 1$) indicating a likely deliberate response.

$\phi^i$ may be considered a signal-to-noise measure for observed demand response. In fact, if $B^i$ is normally distributed then $\phi^i$ is monotonically related to the $z$-statistic
\begin{equation} \label{eq:zDefinition}
	z^i = (\mu_{B^i} - b^i) / \sigma_{B^i}
\end{equation}
by the one-to-one mapping $\phi = \frac{1}{\sqrt{\pi} } \int_{-\infty}^z e^{-x^2}\mathrm{d}x$. Definition \eqref{eq:zDefinition} may be read as a signal to noise ratio, where the ``noise'' $\sigma_{B^i}$ is the standard deviation of the random bill $B^i$, and the ``signal'' the difference between its mean and the actual bill. Although we will use $\phi^i$ in the following sections, the analysis may equivalently be performed using $z^i$. The nonparametric quantile $\phi^i$ may be used for arbitrary distributions, whereas the $z$-statistic may be preferable to distinguish between highly-responsive customers if the normal approximation for $B^i$ is appropriate (which is the case for the LCL trial data). 

In practice, the distribution of the random bill $B_i$ and the values $\phi^i$ and $z^i$ are estimated by repeated random sampling of shuffled price signals (100,000 per household in our examples). Fig.~1a illustrates the process for a single household.

\begin{figure}[!t]
	\centering
	\includegraphics[width=\columnwidth]{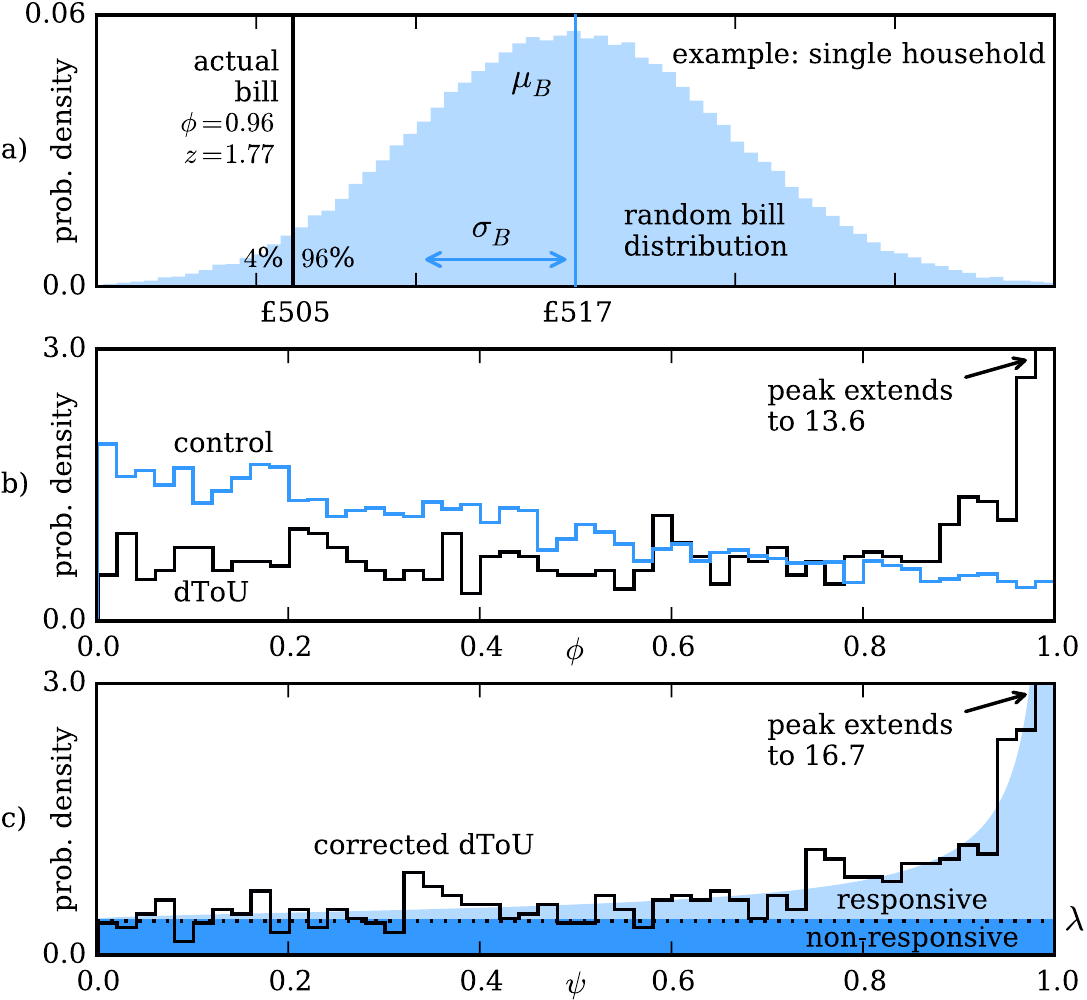}
	\caption{Examples using the LCL trial data. a) Construction of $q$ and $z$ metrics for a single household. b) Distributions of $\{\phi^i\}$ for dToU and control groups (50 bins). c) Corrected dToU distribution $\{\psi^i\}$ and proposed mixture model of subpopulations. Distributions were fitted to $f_\mathrm{dtou}(\psi;\lambda, \alpha, \beta) = \lambda + (1-\lambda) \cdot \mathrm{Beta}(\psi;\alpha, \beta)$, with ${\lambda=0.38}, {\alpha=1.49}, {\beta=0.19}$.}
	\label{fig:distributions}
\end{figure}

\section{Analysing customer responsiveness}
A strong response (high~$\phi^i$ or~$z^i$) can be attributed to:
\begin{itemize}
\item \emph{Deliberate demand response}: a conscious effort by the customer to respond to price signals or autonomous actions by appliances or an energy management system.
\item \emph{Accidental demand response}: coincidental alignment of 
consumption patterns with the price signal. 
\item \emph{Price signal bias}, reflecting an overall dependence between the price signal and the population energy consumption. This may result from deliberate targeting of high load levels using high prices, or from accidental correlation between price signals and demand levels. 
\end{itemize}
For an individual customer it is not possible to disentangle these, but three aggregation methods described below can provide further insight.

\subsection{Confidence ranking}
In absence of an absolute benchmark the first approach is to consider customers' \emph{relative} signal to noise metrics, as captured by the confidence rank 
\begin{equation}
r^i = \textrm{rank index of }\phi^i \textrm{ in }\{ \phi^1, \ldots, \phi^N  \}.
\end{equation}
Customers with a high rank index $r^i$ have outperformed most other customers in terms of their deviation from typical price-agnostic (random) behaviour. The confidence rank correlates strongly with observed demand response in the LCL trials: Spearman's rank correlation with high price response was 0.78 and 0.43 for low price signals (see \cite{Schofield2015} for DR methodology). 

\subsection{Confidence levels}
A second approach is to convert the responsiveness metric $\phi^i$ into a confidence level by comparison with a control group that represents the null hypothesis of non-responsiveness. The control group consists of customers who are assumed to resemble those in the response group, but did not receive the dynamic time-of-use signal. Their $\phi$-values are computed identically. Figure~\ref{fig:distributions}b shows the empirical $\phi$-distributions for LCL trial customers receiving the dToU signal and those on a regular tariff. For the dToU group, the data clearly evidences demand response, with a significant fraction of households (the peak at $\phi=1$) outperforming all 100,000 randomised bills. 

Accidental demand response alone would result in a uniform $\phi$ distribution, so the skewed control group distribution suggests a price signal bias towards increased bills (small $\phi$). This bias can 
be modelled as a coordinate transformation $g: \psi \mapsto \phi$ from the unbiased unit coordinate $\psi$. Let $F_{\mathrm{\mathrm{control}}}(\phi)$ be the (empirical) cumulative probability distribution of $\{\phi^i\}$ in the control group so that it represents the proportion of customers with $\phi^i \le \phi$. In order for the transformation $g$ to result in an unbiased corrected distribution, it must satisfy 
$\hat{F}_{\mathrm{control}}(\psi) \equiv F_{\mathrm{control}}(g(\psi)) = \psi$, which implies $g^{-1}(\phi) := F_\mathrm{control}(\phi)$.
The observed responses $\{\phi^i\}$ of the dToU group can thus be transformed to the bias-corrected responses $\{\psi^i\}$ using
\begin{equation}
	\psi^i = F_\mathrm{control}(\phi^i), \qquad \psi^i \in [0,1].
\end{equation}
Intuitively, this can be understood as a fraction $\psi^i$ of the control group having a response less than or equal to $\phi^i$.  

The bias-corrected $\psi$-distribution for the dToU group is shown in Figure~\ref{fig:distributions}c. With the price signal bias accounted for, the value $\psi^i$ may be used as a direct measure of the level of confidence that a household $i$ responded deliberately. Of the 988 households in the dToU group, 41\% responded deliberately at the 95\% confidence level ($\psi^i \ge 0.95$). 

\subsection{Subpopulation identification}
The third method considers responsiveness at a population level instead of assessing each customer individually. The bias-corrected $\psi$-distribution can be partitioned into a mixture of responsive and non-responsive subpopulations. The presence of a uniform background level $\lambda$ is assumed to arise from the accidental demand response of the unresponsive population; the rest of the population is considered responsive. 

Figure~\ref{fig:distributions}c shows the fit of the proposed distribution $f_{\mathrm{dtou}}(\psi;\lambda,\alpha,\beta)$ to the corrected dToU distribution. The fitted parameter $\lambda=0.38$ suggests that 62\% of dToU customers were responsive, i.e. they demonstrated behaviour that was clearly distinguishable from the control group. Note that this result does not depend on the choice of a confidence level. While it is impossible to state with certainty to which population any individual customer $i$ belongs, we may compute the probability of responsiveness conditional on its computed $\psi^i$ value:
\begin{equation} \label{eq:responsive}
	\mathrm{Pr(responsive}|\psi^i) = 1 - \lambda / f_{\mathrm{dtou}}(\psi^i)
\end{equation}

\section{Conclusion}
We have introduced a novel nonparametric measure of customer responsiveness to dToU tariffs, which may be interpreted as a signal to noise metric for an individual customer's response---without requiring the use of baseline models. Three increasingly sophisticated methods have been described to analyse these measurements to identify responsive customers. The resulting customer classification may be used to tailor information that is sent to customers, or to extrapolate trial results to populations with more or fewer responsive customers.


\begin{thebibliography}{1}
\providecommand{\url}[1]{#1}
\csname url@samestyle\endcsname
\providecommand{\newblock}{\relax}
\providecommand{\bibinfo}[2]{#2}
\providecommand{\BIBentrySTDinterwordspacing}{\spaceskip=0pt\relax}
\providecommand{\BIBentryALTinterwordstretchfactor}{4}
\providecommand{\BIBentryALTinterwordspacing}{\spaceskip=\fontdimen2\font plus
\BIBentryALTinterwordstretchfactor\fontdimen3\font minus
  \fontdimen4\font\relax}
\providecommand{\BIBforeignlanguage}[2]{{%
\expandafter\ifx\csname l@#1\endcsname\relax
\typeout{** WARNING: IEEEtran.bst: No hyphenation pattern has been}%
\typeout{** loaded for the language `#1'. Using the pattern for}%
\typeout{** the default language instead.}%
\else
\language=\csname l@#1\endcsname
\fi
#2}}
\providecommand{\BIBdecl}{\relax}
\BIBdecl

\bibitem{Mathieu2011}
J.~L. Mathieu, D.~S. Callaway, and S.~Kiliccote, ``{Examining uncertainty in
  demand response baseline models and variability in automated responses to
  dynamic pricing},'' in \emph{IEEE Conference on Decision and Control and
  European Control Conference}.\hskip 1em plus 0.5em minus 0.4em\relax IEEE,
  Dec. 2011, pp. 4332--4339.

\bibitem{Wijaya2014}
T.~Wijaya, M.~Vasirani, and K.~Aberer, ``{When Bias Matters: An Economic
  Assessment of Demand Response Baselines for Residential Customers},''
  \emph{IEEE Transactions on Power Systems}, vol.~5, no.~4, pp. 1755--1763,
  2014.

\bibitem{Schofield2015}
J.~R. Schofield, ``{Dynamic time-of-use electricity pricing for residential
  demand response: Design and analysis of the Low Carbon London smart-metering
  trial},'' Ph.D. dissertation, Imperial College London, 2015.

\end{thebibliography}
\end{document}